\documentclass[aps,pra,showpacs,twocolumn]{revtex4}
\usepackage{amssymb}
\usepackage{epsfig}
\usepackage{graphicx}
\usepackage{amsmath}
\usepackage{array,color}

\begin{document}

\title{Exact dark state solutions of the coupled atomic-molecular Bose-Einstein condensates in an external potential}
\author{Xiao-Fei Zhang$^{1,3}$, Jun-Chao Chen$^{2}$, Biao Li$^{2}$, Lin Wen$^{1}$, and W. M. Liu$^{1}$}
\address{$^1$Beijing National Laboratory for Condensed Matter Physics,
Institute of Physics, Chinese Academy of Sciences, Beijing 100190,
China}

\address{$^2$Nonlinear Science Center, Ningbo University, Ningbo 315211,
China}
\address{$^3$College of Science, Honghe University, Mengzi 661100, China}

\date{\today}

\begin{abstract}
We consider a coupled nonlinear Schr\"{o}dinger equations describing
an atomic Bose-Einstein condensates coupled to a molecular
condensates through the stimulated Raman adiabatic passage loaded in
an external potential. The existence of dark state are investigated
within the full parameter space accounts for all the nonlinear
collisions, together with the atom-molecule conversion coupling and
external potential. The results show that there are a class of
external potentials such as double-well, periodical and double
periodical, and harmonic potentials, where the exact dark solutions
can be formed. Our results may raise the possibility of relative
experiments for dark state in the coupled atomic-molecular
Bose-Einstein condensates.
\end{abstract}

\maketitle

\section{Introduction}

The experimental realization of Bose-Einstein condensates (BECs)
\cite{M. H. Anderson,C. C. Bradley,K. B. Davis} arouse great
interest in ultracold molecule due to their potential applications
for the tests of fundamental physics and for the drifts of
fundamental constants \cite{M. Greiner,M. Bartenstein,J. R.
Abo-Shaeer,M. Bartenstein1,J. J. Hudson}. This fundamental system is
of special interest since the collisional coherence can led to
remarkable demonstration of the reversible atom-molecule formation
\cite{P. D. Drummond}; thus open the floodgate for many beautiful
experiments, such as coherent oscillations between atoms and
molecules, dissociation of molecular BECs, the ``clumping" of the
condensates due to the modulational instability, and so on \cite{E.
Timmermans,Ming,S. L. W. Midgley}. It also represents a rich ground
for investigations of the nonlinear excitation of the coupled
atomic-molecular Bose-Einstein condensates (AMBEC) \cite{T. J.
Alexander}.

In real experiments, the cold molecules can be produced from a Fermi
gas of atoms \cite{C. A. Regal,Markus Greiner,J. Cubizolles} or an
atomic BEC based on Feshbach resonances, Raman photoassociation, or
stimulated Raman adiabatic passage (STIRAP) \cite{E. Timmermans1,R.
Wynar,E. A. Donley,B. Damski,R. A. Duine,Volz,G. Thalhammer,T.
Volz,H. Y. Ling}. However, among them, the Feshbach coupling
mechanism is restricted to the creation of molecules in the highest
rovibrational level and is only practicable for a limited number of
systems \cite{K. Winkler,S. Inouye,J. Stenger}, and the losses
caused by inelastic atom-molecule collisions occurs at a significant
rate. In the Raman photoassociation process, the effective
conversion rate of a pair of atoms to ground molecule is limited by
spontaneous emission from the excited molecular state; while the
STIRAP is known to have the highest rate of efficiency when
converting an atomic condensates into a molecular one based on a
Raman transition, where the input Raman laser pulse couples the
molecular levels and reduces spontaneous emission. During this
process, the molecular state is in a dark superposition state (or
coherent population trapping state, CPT state), which decouples from
the light and thus suppresses losses due to spontaneous light
scattering \cite{K. Winkler1}. In this situation, the excited
molecular state of this coupled system remains almost unpopulated.
Thus it is not difficult to see the existence of dark is the key
ingredient in effective production of molecule through STIRAP
technique.

Generally speaking, the dark state cannot exist with any external
potential as the coupled nonlinear Schr\"{o}dinger equations (see
Eq. (\ref{CGP}) below) is nonintegrable \cite{M. Mackie,H. Pu,A. P.
Itin,S. Y. Meng,X. F. Zhang,H. A. Cruz}. The main purpose of the
present paper is to find the exact dark state solutions of this
coupled system with tunable nonlinear interactions and external
potential. Based on the framework of a three-component mean-field
model that takes into account all types of the mean-field nonlinear
atomic and molecular interactions and external potential. we present
a comprehensive analysis of the dark states in the parametrically
coupled atomic-molecular Bose-Einstein condensates. The results show
that there are a class of external potentials, such as double-well,
periodical and double periodical, and harmonic potentials, which can
be used to support the existence of the dark states.

This paper is organized as follows. In sec. II, we briefly outline
the theoretical model under study and the underlying mean-field
coupled equations. Then a class of simplification is presented and
the reduced coupled equations, focusing on the CPT solutions, are
obtained. In Sec. III, the analytical methods for solving the
coupled Schr\"{o}dinger equations are introduced, and we present
several classes of exact dark state solutions where different
potentials can be applied to trap the coupled condensates and
relationship between system parameters are established. A simple
summary is given in the final section.

\section{The theory model}

To begin with, we consider a quasi-one-dimensional geometry of the
condensates where the transverse traps are tight enough, thus the
transverse motion of the condensates are frozen to the ground state
of the transverse harmonic trapping potential. The coupled
Heisenberg equations of motion for the atomic and molecular
components are govern by the system of equations as follows:

\begin{eqnarray}
i\hbar\frac{\partial \psi_{1}}{\partial t} &=&
-\frac{\hbar^2}{2m}\frac{
\partial^{2}\psi_{1}}{\partial x^{2}}+ V_1 \psi_{1} -\frac{\hbar\Omega_1}{\sqrt{2}}%
\psi_3\psi_1^{*} \notag \\
&&+ ( U_{11}|\psi_{1}|^{2}+U_{12}|\psi_{2}|^{2}+
U_{13}|\psi_{3}|^{2})\psi_{1}, \notag
\\
i\hbar\frac{\partial \psi_{2}}{\partial t} &=&
-\frac{\hbar^2}{4m}\frac{
\partial^{2}\psi_{2}}{\partial x^{2}}+ V_2 \psi_{2} -\frac{\hbar\Omega_2}{2}%
\psi_3  \notag \\
&&+ ( U_{21}|\psi_{1}|^{2}+U_{22}|\psi_{2}|^{2}+
U_{23}|\psi_{3}|^{2})\psi_{2}, \notag
\\
i\hbar\frac{\partial \psi_{3}}{\partial t} &=&
-\frac{\hbar^2}{4m}\frac{
\partial^{2}\psi_{3}}{\partial x^{2}}+ V_3 \psi_{3} -\frac{\hbar\Omega_1}{2\sqrt{2}}%
\psi_1^2 -\frac{\hbar\Omega_2}{2}\psi_{2} \notag \\
&&+ ( U_{31}|\psi_{1}|^{2}+U_{32}|\psi_{2}|^{2}+
U_{33}|\psi_{3}|^{2})\psi_{3}, \label{CGP}
\end{eqnarray}
here we define the order parameter $\psi _{a}$ represents an atomic
species of mass $m$ in a potential $V_1(x)$; while $\psi_2$ and
$\psi_3$ represents the ground and excited molecular species with a
potential $ V_{2}(x)$ and $ V_{3}(x)$, respectively. The nonlinear
atom-atom interaction $U_{11}=4\pi\hbar^2 a_{11}/m$, the
atom-molecule interaction $U_{12}=U_{21}=(3a_{12}/4a_{11})U_{11}$,
and the ground molecule-molecule interaction
$U_{22}=(a_{22}/2a_{11})U_{11}$, with $a_{ij}$ being the
corresponding $s$-wave scattering lengths modulated by a Feshbach
resonance. $\Omega_1$ is the Rabi frequency corresponding to the
transitions between the atomic state and the excited molecular
state, and $\Omega_2$ is the one between the molecular ground and
excited states. Finally, in this paper, we will only focus on the
degenerate case with $V_2=V_3=2V_1$ and ignore loss and growth
mechanisms \cite{H. A. Cruz,G. Orso}; thus the total number of
atomic particles, including pairs of atoms inside the diatomic
molecules is conserved in this model.

From a general mathematical and nonlinear physical point of view,
Eq. (\ref{CGP}) is an example of a classically nonintegrable field
theory, which needs to be treated numerically. To obtain the exact
dark state solutions, we make a class of simplifications. To this
end, we first neglect the excited molecular state as it has a small
population compared with atomic and ground molecular states, and
then set $U_{3i}=0$. Similar to homogeneous condensates cases, the
STIRAP model considered in this paper is also found to support a CPT
state (with $\psi_3(x,t)=0$) in the form of

\begin{equation}
\psi_1(x,t)=\phi_1(x)e^{-i\mu t}, \psi_2(x,t)=\phi_2(x)e^{-i 2\mu
t}. \label{che}
\end{equation}

Normalizing the time and length in Eq. (\ref{CGP}) by $t'=\omega_2
t/2 $ and $x'=\sqrt{\omega m/\hbar}x$ (the tilde is omitted for
simplicity in the following discussions). Then inserting this
equation into Eq. (\ref{CGP}), one can readily derive the following
stationary equations:

\begin{eqnarray}
\mu \phi_1=-\frac{d^2 \phi_1}{d x^2} + (g_{11} \phi_1^2 +
g_{12} \phi_2^2) \phi_1+V(x)\phi_1,  \notag \\
\mu \phi_2=-\frac{1}{4}\frac{d^2 \phi_2}{d x^2} + \frac{1}{2}(g_{22}
\phi_2^2 + g_{21} \phi_1^2) \phi_2+ V(x)\phi_2, \label{CGP1}
\end{eqnarray}
with $\phi_2=-\Omega \phi_1^2 /2$. Here we use the fact
$\mu_2=2\mu_1=2\mu$ and the definition $\Omega=\Omega_1 / \Omega_2$.
In Ref. \cite{H. A. Cruz}, Eq. (\ref{CGP1}) is reduced to a single
ordinary differential equation and the possibilities of existence of
inhomogeneous dark states of atomic-molecular Bose-Einstein
condensates loaded in special external potential is discussed. In
what follows we find that the exact dark state solutions can also
exist in other types of external potentials, such as double-well,
periodical and double periodical potentials.

\begin{figure}[tbp]
\centering
\includegraphics[width=9.5cm]{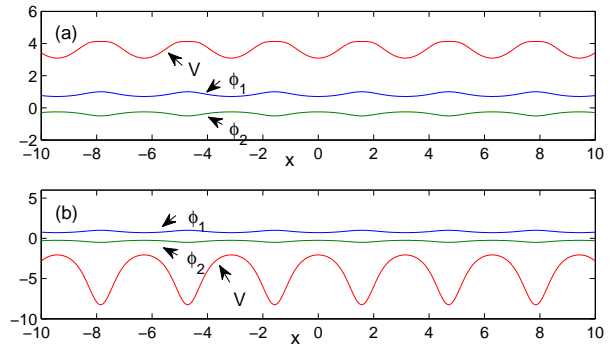}\hspace{4.5cm}
\caption{(Color online) The stationary dark state wave functions of
the atomic and molecular condensates $\phi_{11}(x)$ and
$\phi_{21}(x)$ given by Eq. (\ref{ES11}), and the profile of the
external potential in the limit $k=0$ and $a_0=0$. The parameters
are given as follows: $\delta=1$,$a_0=0$,
$B_1=a_1=A_0=\Omega=\lambda=\mu=1$, and $g_{22}=1$ in Fig. 1a and
$g_{22}= 34$ in Fig. 1b. As shown in this figure, different external
potentials but with similar periodical structure can support the
same stationary dark state wave functions of the atomic and
molecular condensates.} \label{F1}
\end{figure}

\section{Exact dark state solutions and discussion}

In this section, we introduce a analytical method to derive a class
of exact dark state solutions for Eq. (\ref{CGP1}). Without a trap,
i.e., $V_i=0$, the system is homogeneous and the dark state is well
studied (see Refs. \cite{M. Mackie,H. Pu,H. Y. Ling1} and references
therein). In the presence of external potential, we find that some
special types of dark state solutions can be derived via suitable
selections of the nonlinear parameters and external potential.

To get the analytical dark state solutions, we assume the stationary
dark state solutions have the following forms:

\begin{eqnarray}
\phi_1(x) &=& \delta \sqrt{a_0+a_1 \varphi}, \notag\\
\phi_2(x)&=&-\frac{1}{2}\Omega(a_0+a_1\varphi), \label{S}
\end{eqnarray}
with auxiliary function $\varphi=\varphi(x)$ satisfying the
following equations

\begin{eqnarray}
(\frac{d\varphi }{d x})^2&=&h_0+h_1\varphi+h_2
\varphi^2+h_3 \varphi^3+h_4\varphi^4, \notag\\
\frac{d^2\varphi}{d x^2}&=& \frac{h_1}{2} + h_2 \varphi
+\frac{3}{2}h_3\varphi^2 +2 h_4 \varphi^3, \label{Au}
\end{eqnarray}
where $\delta=\pm1$ and $a_0, a_1, h_0, h_1, h_2, h_3, h_4$ are
constants to be determined. Now we shift our attention to the
external potential and further assume the external potential is in
the form of

\begin{equation}
V=\frac{b_0+b_1\varphi+b_2 \varphi^2+b_3 \varphi^3+b_4
\varphi^4}{d_0+ d_1 \varphi}, \label{V}
\end{equation}
with $b_0, b_1, b_2, b_3, b_4, d_0,d_1$ are constants to be
determined. Substituting Eqs. (\ref{S}) and (\ref{V}) into Eq.
(\ref{CGP1}) and using the auxiliary Eq. (\ref{Au}), we can obtain a
set of ordinary differential equations (ODEs) with respect to the
controllable system parameters \{$\Omega, g_{11}, g_{12}, g_{22}$\}.
As shown in Eqs. (\ref{S}) and (\ref{Au}), depending on different
choices of the form of $\varphi(x)$, we can get various types of
external potential and exact dark state solutions through Eqs.
(\ref{S}) and (\ref{V}). It should be mentioned that in order to
simplify the calculation, we further set $\delta=1$ in the following
discussions. Finally, solving these ODEs, we can obtain two families
of analytical solutions of Eq. (\ref{CGP1}).

\begin{figure}[tbp]
\centering
\includegraphics[width=9.0cm]{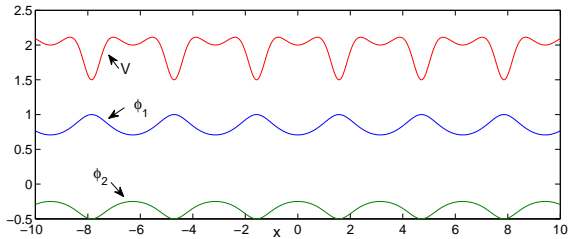}\hspace{4.0cm}
\caption{(Color online) The stationary dark state wave functions of
the atomic and molecular condensates $\phi_{11}(x)$ and
$\phi_{21}(x)$ given by Eq. (\ref{ES11}), and the profile of
external potential in the limit $k=0$ and $a_0=B_0=0$. The other
parameters are the same as in Fig (\ref{F1}) except $g_{22}=8$. In
this case, the shape of external potential shows double-periodical
structure.} \label{F2}
\end{figure}

\subsection{The Jacobi cn-wave solutions}

We begin with the Jacobi cn-function solution of $\varphi(x)$. In
this case, we can get

\begin{equation}
\varphi_1= \frac{A_0}{B_0+B_1 {\rm cn}^2(\lambda x, k)},
\end{equation}
where {\rm cn} is the Jacobi elliptic function and $0\leq k\leq 1$
is the module of Jacobi elliptic functions, $B_0\neq -B_1$. At this
point, the analytical dark state solutions for the atomic and
molecular condensates can be written as

\begin{eqnarray}
\phi_1=\sqrt{\frac{a_1 A_0 }{ B_0+B_1 {\rm cn}^2(\lambda x, k)}}, \notag\\
\phi_2= \frac{-\Omega a_1 A_0}{2B_0+2B_1 {\rm cn}^2(\lambda x, k)},
\label{ES1}
\end{eqnarray}
here we set  $a_0=0$. In this situation, the external potential can
be obtained by solving Eq. (\ref{V}) and reads

\begin{equation}
V_1(x) = (\frac{1}{2}\Omega^2 g_{12} -\frac{3}{8}\Omega^2
g_{22})\varphi_1^2(x) + (3 g_{11}-2 g_{12})\varphi_1(x)+\mu+
\frac{h_2}{4}. \label{EP1}
\end{equation}

It is evident that the external potential is a function of nonlinear
parameters, chemical potential, the Rabi frequencies, and the
auxiliary function $\varphi(x)$. In the following discussions, we
find that with different choices of the auxiliary function
$\varphi(x)$, there exist a class of external potentials, such as
double-well, periodical and double periodical, and harmonic
potentials, where analytical dark state can exist and the constraint
conditions for the nonlinear parameters can be obtained. For
simplicity, we only discuss the exact dark state solutions of Eq.
(\ref{CGP1}) in two cases with the module of Jacobi elliptic
functions $k=0$ and $k=1$.

\textit{Case 1}. In the limit $k=0$, one can easily get
\begin{eqnarray}
\phi_{11}=\sqrt{\frac{a_1 A_0 }{ B_0+B_1\cos^2(\lambda x)}}, \notag\\
\phi_{21}= \frac{-\Omega a_1 A_0}{2B_0+2B_1 \cos^2(\lambda x)}.
\label{ES11}
\end{eqnarray}
where $\lambda=\pm1$. In this case, the relationship between
interaction parameters read

\begin{eqnarray}
&&g_{12}= \frac{-4 \lambda^2(B_1B_0 +B_0^2)}{\Omega^2 a_1^2} + \frac{g_{22}}{2}, \notag\\
&&g_{11}= \frac{-2 \lambda^2(B_1B_0 +B_0^2)}{\Omega^2 a_1^2} +
\frac{\lambda^2B_0}{a_1}+ \frac{B_1\lambda^2}{2a_1} +
\frac{g_{22}}{4}, \label{g0}
\end{eqnarray}
with $A_0=1$ and $g_{22}$ a function of the system parameters.

\begin{figure}[tbp]
\centering
\includegraphics[width=9.0cm]{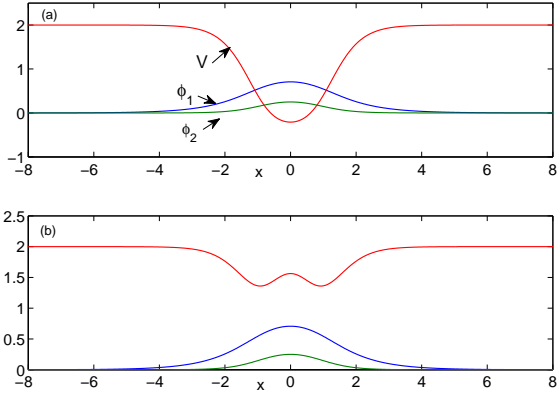}\hspace{4.5cm}
\caption{(Color online) The stationary dark state wave functions of
the atomic and molecular condensates $\phi_{12}(x)$ and
$\phi_{22}(x)$ given by Eq. (\ref{ES12}), and the profile of
external potential in the limit $k=1$. The parameters are given as
follows: $\delta=1$, $B_0= B_1=a_0=1$, $\lambda=1$, $\Omega=-1$, and
$g_{22}=-20/3$ in (a) and $g_{22}=-18 $ in (b).} \label{F3}
\end{figure}

Shown in Fig. \ref{F1} are the profiles of the wave functions for
the atomic and molecular condensates $\phi_1(x)$ and $\phi_2(x)$,
and the external potential in the limit $k=0$. The parameters are
given as $\delta=1$, $B_1=a_1=A_0=\Omega=1$, $\lambda=1$, and
$g_{22}=1$ in Fig. 1(a) and $g_{22}= 34$ in Fig. 1(b). According to
Eq. (\ref{g0}), which gives the relationship among these three kinds
of nonlinear interactions; for Fig. 1(a), both the atom-atom and
atom-molecule interactions $g_{11}=-2.25<0$ and $g_{12}=-7.5 < 0$,
corresponding to attractive atom-atom and atom-molecule
interactions; while for Fig. 1(b), $g_{11}>0$ and $g_{12} > 0$,
which together with $g_{22}= 34$, corresponding to repulsive
atom-atom, atom-molecular, and molecule-molecule interactions.

As shown in this figure, different external potentials but with
similar periodical structure can support the same stationary dark
state solution. However, we want to point out the difference between
Fig. \ref{F1}(a) and Fig. \ref{F1}(b). For Fig. \ref{F1}(a), the
external potential is relative shallow and the density distribution
is mainly locating at the crest, which is the result of the
competition between the nonlinear interaction and the external
potential; while for Fig. \ref{F1}(b), when the atom-atom and
atom-molecule interaction become repulsive, the height of the
external potential increases, and the density distributions of the
order parameters are different from the former one and mainly
locating at the wave trough.

Figure \ref{F2} shows the profiles of the atomic and molecular
condensates $\phi_1(x)$ and $\phi_2(x)$ and external potential in
the case of $g_{22}=8$. In this case, $g_{11}=-0.5$ and $g_{12}=-4$,
which are smaller than the parameters used in Fig. 1(a), but with a
stronger molecule-molecule interactions. It is easy to see that the
external potential presented here shows double-periodical structure,
which can be realized by two pairs of counter-propagation lasers,
and widely used in today's ultracold atom experiments. Finally, no
matter the periodical or the double-periodical potentials, the
density distribution always continuous distribution within the
extension of the external potential. This is different from the
localized solutions reported in \cite{H. A. Cruz} and the following
solutions, where the chemical potential must be smaller than zero.

Thus we conclude that when the ratio of the Rabi frequency is fixed,
both periodical and double-periodical potentials can be applied to
the coupled atomic-molecular Bose-Einstein condensates to get the
analytical dark state solution via varying the strength of the
nonlinear interactions.

\begin{figure}[tbp]
\centering
\includegraphics[width=8.0cm]{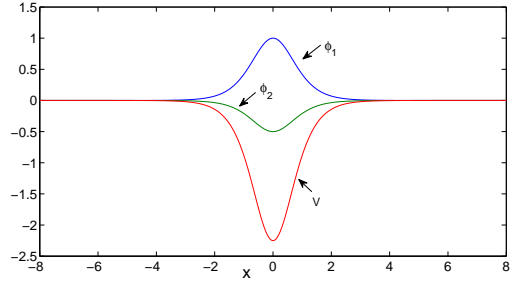}
\caption{(Color online) The stationary dark state wave functions of
the atomic and molecular condensates $\phi_{13}(x)$ and
$\phi_{23}(x)$ given by Eq. (\ref{ES2}), and the profile of the
external potential given by Eq. (\ref{V2}). The parameters are given
as follows: $\delta=1,\lambda=1, a_0=0, \Omega=1, A_0=1, g_{12}=1,
a_1=1, g_{22}=20,\mu=-1$.} \label{F4}
\end{figure}

\textit{Case 2}. Now we shift our attention to the case of $k=1$. In
this situation, the stationary solutions for the atomic and
molecular condensates can be written as

\begin{eqnarray}
\phi_{12}=\sqrt{\frac{a_0 B_1{\rm sech}^2(\lambda x) }{ B_0+B_1 {\rm
sech}^2(\lambda x)}}, \notag\\
\phi_{22}= \frac{-\Omega a_0 B_1 {\rm sech}^2(\lambda x)}{2B_0+2B_1
{\rm sech}^2(\lambda x)}. \label{ES12}
\end{eqnarray}

The relationships among the interaction parameters in this case read

\begin{eqnarray}
&&g_{12}=  \frac{4\lambda^2(B_0 +B_1)}{\Omega^2 a_0^2 B_1} + g_{22}/2, \notag\\
&&g_{11}=\frac{2 \lambda^2(B_0 +B_1)}{\Omega^2 a_0^2 B_1}
-\frac{\lambda^2}{a_0} -\frac{B_0\lambda^2}{2a_0
B_1}+\frac{g_{22}}{4},
\end{eqnarray}
with $g_{22}$ satisfying the following two cases

\begin{eqnarray}
&&g_{22}^{i}=\frac{4a_0B_0 \lambda^2
\Omega^2-16\lambda^2B_0-16B_1\lambda^2+8a_0 \Omega^2 B_1}{B_1 a_0^2
\Omega^2 (2+\Omega^2 a_0)},
\notag\\
&&g_{22}^{ii}=
\frac{-8\lambda^2(B_0+B_1)}{a_0^2\Omega^2B_1}-\frac{4B_1\lambda^2+
6B_0\lambda^2}{a_0 B_1} + \frac{8}{a_0}. \label{g22}
\end{eqnarray}

Shown in Fig. \ref{F3} are the wave functions of the atomic and
molecular condensates $\phi_{12}(x)$ and $\phi_{22}(x)$, and the
profile of external potential in the case of attractive
molecule-molecule interaction. The parameters are given as follows:
$\delta=1$, $B_0= B_1=a_0=1$, $\lambda=1$, $\Omega=-1$. The
molecule-molecule interaction $g_{22}=-20/3$ for Fig. \ref{F3}(a)
and $g_{22}=-18$ for Fig. \ref{F3}(b), corresponding to the two
different cases (i) and (ii) in Eq. (\ref{g22}), respectively.
According to the constrain conditions for the nonlinear parameters,
we can get $g_{11}=5/6$, $g_{12}=14/3$, and $g_{22}=-20/3$ for Fig.
\ref{F3}(a), which corresponding to repulsive atom-atom and
atom-molecular interactions and attractive molecule-molecule
interaction; while for Fig. \ref{F3}(b), the nonlinear parameters
read $g_{11}=-2$, $g_{12}=-1$, and $g_{22}=-18$, where all the
interactions are attractive. As shown in Fig. \ref{F3}(a), the
external potential shows the harmonic structure, while double-well
structure in Fig. \ref{F3}(b), which is similar to the situation in
Fig. (\ref{F1}): the order parameters $\phi_{12}(x)$ and
$\phi_{22}(x)$ are with the same distributions in different
potentials.

It is necessary to point out the stationary dark state solutions
obtained in this case is different from the ones in \textit{Case 1}
as $\phi_{12}(x)$ and $\phi_{22}(x)$ are localized and symmetric
with respect to certain point X (here we choose X=0). Thus we
conclude that the exact localized dark state solutions can be
achieved with a proper choice of the nonlinear interaction
parameters and external potential, and  both harmonic and
double-well potentials can be used to trap the coupled
atomic-molecular Bose-Einstein condensates to get the analytical
dark state solution via varying the strength of the nonlinear
interactions.

\begin{figure}[tbp]
\centering
\includegraphics[width=7.0cm]{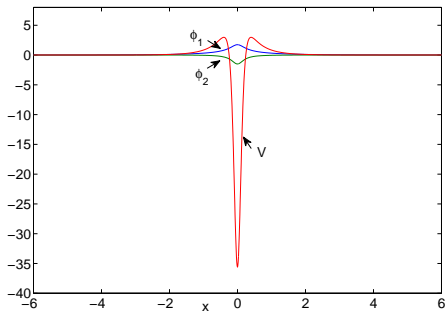}
\caption{(Color online) The same as in Fig. \ref{F4} but for
$a_1=3$.}\label{F5}
\end{figure}

\subsection{The Jacobi sn- and dn-wave solutions}

In this subsection, we consider the Jacobi sn- and dn- function
solution of $\varphi(x)$. In this case, one can easily get

\begin{equation}
\varphi_2= A_0 + \frac{A_1  {\rm sn}(\lambda x, k)^2}{ {\rm
dn}(\lambda x, k)^2+A_3},
\end{equation}
where sn and dn are the Jacobi elliptic functions and $0\leq k\leq
1$ is the module of Jacobi elliptic functions. With the same
procedure in subsection A, we can obtain a family of exact dark
state solutions. As the expressions of wavefunctions for atomic and
molecular condensates are too complicated, here we only focus on the
limit $k=1$ and list some main results.

In the case of $\delta=\lambda=\Omega=A_0=g_{12}=1$ ,
$a_0=0$,$a_1=1$, $g_{22}=20$, $\mu=-1$, the wave functions of the
atomic and molecular condensates can be expressed as

\begin{equation}
\phi_{13}= \frac{csgn (1/\cosh(x))}{\cosh(x)}, \phi_{23}=
\frac{-1}{2\cosh(x)^2}, \label{ES2}
\end{equation}
with csgn the sign function for real and complex expressions. In
this situation, the external potential is in the following form

\begin{equation}
V_2(x)=\frac{-1-8 \cosh(x)^2}{4 \cosh(x)^4}. \label{V2}
\end{equation}

Figure \ref{F4} exhibits the profiles of the order parameters of the
atomic and molecular condensates $\phi_{13}(x)$ and $\phi_{23}(x)$
and the external potential. As shown in this figure, the external
potential is similar to the usual harmonic one and the result is
similar to the one shown in Fig. (1c) in \cite{H. A. Cruz}; while
Fig. \ref{F5} shows the same wave functions with the same parameters
used in Fig. \ref{F4} but for $a_1=3$. It is easy to see that the
external potential in Fig. \ref{F5} is different form the usual
harmonic one, but with a peak in each side of the potential. It
should be mentioned that in this case the nonlinear parameters are
different. Thus we conclude that by manipulation the nonlinear
parameters, there exist a class of external potentials where
analytical dark solution can be obtained. Finally, we want to point
out that the localized stationary dark state solutions and the
corresponding two types of external potentials only exist for $\mu<
0$, which agrees well with the results obtained in \cite{H. A.
Cruz}.

The stability of the localized stationary solutions obtained in this
paper is still an open problem. This can be achieved by carrying out
numerical simulations of the localized stationary solutions with the
imposed initial perturbations, or equivalently adopting the linear
stability analysis: a small deviation away from the intended state
is added to see the dynamical evolution of the system; if the
deviation remains small, then the stability (or adiabaticity) is
obeyed. Since the object of this work is to obtain some exact
analytical solutions, a thorough analysis of the stability is beyond
the scope of the present paper and these works belong to further
studies.

\section{Conclusions}

In summary, we focus on an atomic Bose-Einstein condensates coupled
to a molecular condensates through the stimulated Raman adiabatic
passage loaded in an external potential, with the emphasis on the
stationary dark state solutions. Within the full parameter space
accounts for atom-atom, atom-molecule and molecule-molecule
collisions, together with the atom-molecule conversion coupling and
the external potential, we detailed investigate under which
conditions the exact dark state solution can exist. Our results show
that via suitable selections of the nonlinear parameter and external
potential, the analytical dark state solutions can be constructed
within a class of external potentials including double-well,
periodical and double periodical, and harmonic potentials. The
obtained results are of particular significance to experimental
realization of such dark state in coupled atomic-molecular
Bose-Einstein condensates.

\section{Acknowledgments}

We would like to express our sincere thanks to Professor G. V.
Shlyapnikov for helpful discussions. This work was supported by NSFC
under grants Nos. 10874235, 10934010, 60978019, the NKBRSFC under
grants Nos. 2009CB930701, 2010CB922904, 2011CB921502, 2012CB821300,
and NSFC-RGC under grants Nos. 11061160490 and 1386-N-HKU748/10.


\begin{thebibliography}{99}

\bibitem{M. H. Anderson} M. H. Anderson, J. R. Ensher, M. R. Matthews, C. E. Wieman and
E. A. Cornell, Science {\bf 269}, 198 (1995).

\bibitem{C. C. Bradley} C. C. Bradley, C. A. Sackett, J. J. Tollett, and R. G. Hulet,
Phys. Rev. Lett. {\bf 75}, 1687 (1995).

\bibitem{K. B. Davis} K. B. Davis, M. -O. Mewes, M. R. Andrews, N. J. van Druten, D.
S. Durfee, D. M. Kurn, and W. Ketterle, Phys. Rev. Lett. {\bf 75},
3969 (1995).

\bibitem{M. Greiner} M. Greiner, C. A. Regal, and D. S. Jin, Nature {\bf 426}, 540 (2003).

\bibitem{M. Bartenstein} M. Bartenstein, A. Altmeyer, S. Riedl, S. Jochim, C. Chin, J. H.
Denschlag, and R. Grimm, Phys. Rev. Lett. {\bf 92}, 120401 (2004).

\bibitem{J. R. Abo-Shaeer} J. R. Abo-Shaeer, D. E. Miller, J.K. Chin, K. Xu, T. Mukaiyama,
andW. Ketterle, Phys. Rev. Lett. {\bf 94}, 040405 (2005).

\bibitem{M. Bartenstein1} M. Bartenstein, A. Altmeyer, S. Riedl, R. Geursen, S. Jochim, C.
Chin, J. Hecker Denschlag, R. Grimm, A. Simoni, E. Tiesinga, C. J.
Williams, and P. S. Julienne, Phys. Rev. Lett. {\bf 94}, 103201
(2005).

\bibitem{J. J. Hudson} J. J. Hudson, B. E. Sauer, M. R. Tarbutt, and E. A. Hinds, Phys.
Rev. Lett. {\bf 89}, 023003 (2002).

\bibitem{P. D. Drummond} P. D. Drummond, K. V. Kheruntsyan, and H. He, Phys.
Rev. Lett. {\bf 81}, 3055 (1998).

\bibitem{E. Timmermans} E. Timmermans, P. Tommasini, R. C$\hat{o}$t$\acute{e%
}$, M. Hussein, and A. Kerman, Phys. Rev. Lett. {\bf 83}, 2691
(1999).

\bibitem{Ming} M. S. Chang, Q. S. Qin, W. X. Zhang, L. You, and M. S.
Chapman, Nature Phys. {\bf 1}, 111 (2005).

\bibitem{S. L. W. Midgley} S. L. W. Midgley, S. W\"{u}ster, M. K. Olsen, M.
J. Davis, and K. V. Kheruntsyan, Phys. Rev. A \textbf{79}, 053632 (2009).

\bibitem{T. J. Alexander} T. J. Alexander, E. A. Ostrovskaya, Y. S. Kivshar,
and P. S. Julienne, J. Opt. B {\bf 4}, S33 (2002).

\bibitem{Markus Greiner} M. Greiner, C. A. Regal, and D. S. Jin,
Nature (London) {\bf 426},  537 (2003).

\bibitem{C. A. Regal} C. A. Regal, C. Ticknor, J. L. Bohn, and D. S. Jin,
Nature (London) {\bf 424}, 47 (2003).

\bibitem{J. Cubizolles} J. Cubizolles, T. Bourdel, S. J. J. M. F. Kokkelmans,
G. V. Shlyapnikov, and C. Salomon, Phys. Rev. Lett. {\bf 91}, 240401
(2003).

\bibitem{E. Timmermans1} E. Timmermans, P. Tommasini, M. Hussein, and A.
Kerman, Phys. Rep. {\bf 315}, 199 (1999).

\bibitem{R. Wynar} R. Wynar, R. S. Freeland, D. J. Han, and D. J. Heinzen,
Science {\bf 287}, 1016 (2000).

\bibitem{E. A. Donley} E. A. Donley, N. R. Claussen, S. T. Thomson,
and C. E. Wieman, Nature (London) {\bf 417}, 529 (2002).

\bibitem{B. Damski} B. Damski, L. Santos, E. Tiemann, M. Lewenstein, S.
Kotochigova, P. Julienne, and P. Zoller, Phys. Rev. Lett. {\bf 90},
110401 (2003).

\bibitem{R. A. Duine} R. A. Duine and H. T. C. Stoof, Phys. Rep. {\bf 396}, 115 (2004).

\bibitem{Volz} S. D$\ddot{u}$rr, T. Volz, A. Marte, and G. Rempe , Phys. Rev.
Lett. {\bf 92}, 020406 (2004).

\bibitem{H. Y. Ling} H. Y. Ling, H. Pu, and B. Seaman, Phys. Rev. Lett. {\bf 93}, 250403
(2004).

\bibitem{G. Thalhammer} G. Thalhammer, K. Winkler, F. Lang, S. Schmid, R.
Grimm, and J. H. Denschlag, Phys. Rev. Lett. {\bf 96}, 050402
(2006).

\bibitem{T. Volz} T. Volz, N. Syassen, D. M. Bauer, E. Hansis, S. D\"{u}rr,
and G. Rempe, Nat. Phys. {\bf 2}, 692 (2006).

\bibitem{K. Winkler} K. Winkler, G. Thalhammer, M. Theis, H. Ritsch,
R. Grimm, and J. Hecker Denschlag, Phys. Rev. Lett. \textbf{95},
063202 (2005).

\bibitem{S. Inouye} S. Inouye, M. R. Andrews, J. Stenger, H.-J. Miesner, D. M.
Stamper-Kurn, and W. Ketterle, Nature (London) {\bf 392}, 151
(1998).

\bibitem{J. Stenger} J. Stenger, S. Inouye, M. R. Andrews, H.-J. Miesner, D. M.
Stamper-Kurn, andW. Ketterle, Phys. Rev. Lett. {\bf 82}, 2422
(1999).

\bibitem{K. Winkler1}K. Winkler, F. Lang, G. Thalhammer, P. V. D. Straten,
R. Grimm, and J. H. Denschlag, Phys. Rev. Lett. {\bf 98}, 043201
(2007).

\bibitem{M. Mackie} M. Mackie, R. Kowalski, and J. Javanainen, Phys. Rev. Lett. {\bf 84}, 3803
(2000).

\bibitem{H. Pu} H. Pu, P.Maenner,W. Zhang, and H. Y. Ling, Phys. Rev. Lett. {\bf 98}, 050406
(2007).

\bibitem{A. P. Itin} A. P. Itin and S. Watanabe, Phys. Rev. Lett. {\bf 99}, 223903 (2007).

\bibitem{S. Y. Meng} S. Y. Meng, L. B. Fu, and J. Liu, Phys. Rev. A {\bf 78}, 053410 (2008).

\bibitem{X. F. Zhang} X. F. Zhang, X. H. Hu, X. X. Liu, and W. M. Liu, Phys. Rev. A {\bf 79}, 033630 (2009).

\bibitem{H. A. Cruz} H. A. Cruz and V. V. Konotop, Phys. Rev. A {\bf 83}, 033603 (2011).

\bibitem{G. Orso} G. Orso, L. P. Pitaevskii, S. Stringari, and M. Wouters, Phys. Rev.
Lett. {\bf 95}, 060402 (2005).

\bibitem{H. Y. Ling1} H. Y. Ling, P. Maenner, W. Zhang, and H. Pu, Phys. Rev. A {\bf 75}, 033615
(2007).

\end{thebibliography}
\end{document}